# Nonlinear Broadband THz Emission from Symmetry Broken Nitrogen–Vacancy Centers in Diamond Crystals

Hani Barhum[1,2*], Cormac McDonnell[1], Tamara Amro[3], Ilya Simanovsky[1], Pavel Ginzburg[1], Nir Bar-Gill[3], Aharon Blank[4], Mohammad Attrash[1,2]

[1] Department of Physical Electronics, Tel Aviv University, Ramat Aviv, Tel Aviv 69978, Israel

[2] Triangle Research and Development Center, Kfar Qara' 3007500, Israel

[3] The Racah Institute of Physics, The Hebrew University of Jerusalem, Jerusalem 91904, Israel

[4] Schulich Faculty of Chemistry, Technion–Israel Institute of Technology, Haifa, Israel

*Corresponding author E-mail: hani.barhom@gmail.com

**Abstract**

Diamond single crystals are promising nonlinear THz sources due to their high damage threshold, transparency, and small dispersion linear dispersion over THz-NIR which enables relaxing the need for additional phase-matching engineering. However, the centrosymmetry of a diamond's lattice prohibits even-order nonlinear effects, including second harmonic generation and optical rectification. We demonstrate broadband THz emission via optical rectification in an NV-doped diamond, where NV centers break inversion symmetry and induce a nonlinear susceptibility in the lattice. THz time-domain spectroscopy reveals single-cycle emission spanning over 4 THz bandwidth, enabled by a high NV density (~200 ppm) and lattice strain. Density functional theory (DFT) confirms the emergence of finite second-order nonlinear susceptibility, directly linking symmetry breaking to THz generation. The wide bandgap and defect-induced strain support efficient THz emission without crystal damage, establishing NV-diamond as a robust platform for high-field ultrabroadband THz generation.

Negatively charged nitrogen vacancy (NV[-]) centers in diamonds are considered an attractive platform for many applications, particularly in the field of quantum technologies.[1–3] This is driven by their unique properties, such as a long spin coherence time at room temperature and a spin dependent photoluminescence. These properties have resulted in the application of NV centers to areas such as magnetic,[4–6] electric[7] and temperature sensing.[8–11] The NV[-] spin can also be manipulated through optical detection of magnetic resonance (ODMR).[12] Diamond itself has been shown to be an attractive material in the field of optics due its favorable optical and mechanical properties, including a high transparency across a broad wavelength range and an extremely high damage threshold. However nonlinear generation is restricted to odd-order nonlinear effects due to its centrosymmetric structure.[13,14] In particular, second-order processes are forbidden. Nonlinearities in diamonds have been studies using novel schemes such as second harmonic (SH) emission due to the presence of grain boundaries,[15] SH emission due to an overlapping symmetry breaking THz field[16] and Raman-resonance enhanced four wave mixing.[17] Recently, NV centers in diamond irradiated with ultrashort pulses below the bandgap were shown to induce a dipole and quadrupole in the crystal lattice, resulting in the emission of second harmonic fields from a surface region.[18] In the THz frequency range, NV centers in diamond were shown to emit single frequency coherent THz photons with the application of a high strength magnetic field.[19] As such, NV centers may also show promise for broadband THz generation through optical rectification of an ultrashort pump pulse. This can be coupled with the potential for NIR-THz phase matching due to the flat spectral response of diamond from the UV to THz bands. Broadband THz generation itself has been investigated in many material platforms including inorganic crystals,[20,21] organic crystals,[22,23] plasmonic metasurfaces[24] and spintronic emitters[25] amongst others. Each of these approaches carries trade-offs between THz bandwidth, conversion efficiency, and material robustness. For example, lithium niobate can produce high-field THz pulses but with narrower bandwidth due to intrinsic phonon absorption, whereas organic crystals offer broad spectra but suffer from lower damage thresholds. NV-diamond, by contrast, offers the prospect of ultrabroad bandwidth combined with the exceptional durability of diamond, making it a compelling new approach for THz generation.

Here we show for the first time emission of broadband THz pulses from NV doped diamond due to optical rectification of a below bandgap NIR ultrashort pulse in the crystal. The results show a new platform with the potential for high field strength THz generation. In the following, we present a comprehensive investigation that combines experimental THz time-domain measurements with first-principles calculations. The theoretical framework, including derivations of the relevant nonlinear optical coefficients and phase-matching conditions for THz generation in NV[-] diamond. The experimental results include the properties of the NV[-]diamond crystal, the observed THz emission spectra, bandwidth and scaling, and the polarization dependencies, along with comparative analysis against conventional THz sources. We also integrate DFT analysis of the NV[-] defect structure to directly support our conclusions by quantifying symmetry breaking and its impact on the nonlinear response. Finally, we discuss the implications of our findings, evaluate the limitations of the establishing NV[-] diamond as a competitive THz source.

Diamond crystals have an exceptional physical and optical properties. Despite these advantageous properties, its intrinsic centrosymmetric crystal structure cannot adapt to second-order nonlinear optical processes, such as optical rectification from occurring naturally. However, NV centers added into diamond locally break this symmetry, thereby enabling nonlinear optical processes. The optical and structural properties of NV-doped diamond are crucial to effectively explore its potential for nonlinear broadband THz emission. Here, we first characterize the NV-diamond sample to establish the link between the observed nonlinear optical effects and the presence of NV centers and associated lattice strain. **Figure 1** shows several optical properties of the diamond sample. A 3D scan of the diamond sample was done using a confocal microscope (Leica microsystems) and is shown in Figure 1a. The scan shows the $NV^-$ centers distributed throughout the diamond sample (see Methods for NV fabrication details). The general crystal dimensions consist of lateral faces in the mm range, with a thickness of ~ 1 mm. The photoluminescence spectrum under 552 nm laser excitation of the sample is shown in Figure 1b. It shows a broad emission centered on approximately 675 nm. The dashed lines indicate the $NV^-$ and $NV^0$ zero phonon lines at 637 and 575 nm respectively, which agrees with previous results in the literature[26]. Figure 1c shows the Raman spectrum with a single peak observed at 1330 cm$^{-1}$, indicating the purity of the sample. No evidence of non-diamond carbon (i.e., G band 1500–1580 cm$^{-1}$)[27] or NV0 (1406 cm$^{-1}$)[28] peaks were observed. Furthermore, the high purity and crystalline quality are essential for effective propagation, and generation of THz waves. It also confirms that no other THz sources contribute. The ODMR spectrum, shown in Figure 1d reveals a resonance peak at 2.87 GHz, with peak splitting of 10 MHz. This splitting indicates significant internal strain within the crystal, originating from lattice distortion around NV centers. The NV centers are embedded parallel to the 111 planes of the crystal.

Therefore, the crystal characteristics support the claim that its purity and internal symmetry breaking will enhance nonlinear susceptibility, further enabling THz generation through the effect of optical rectification.

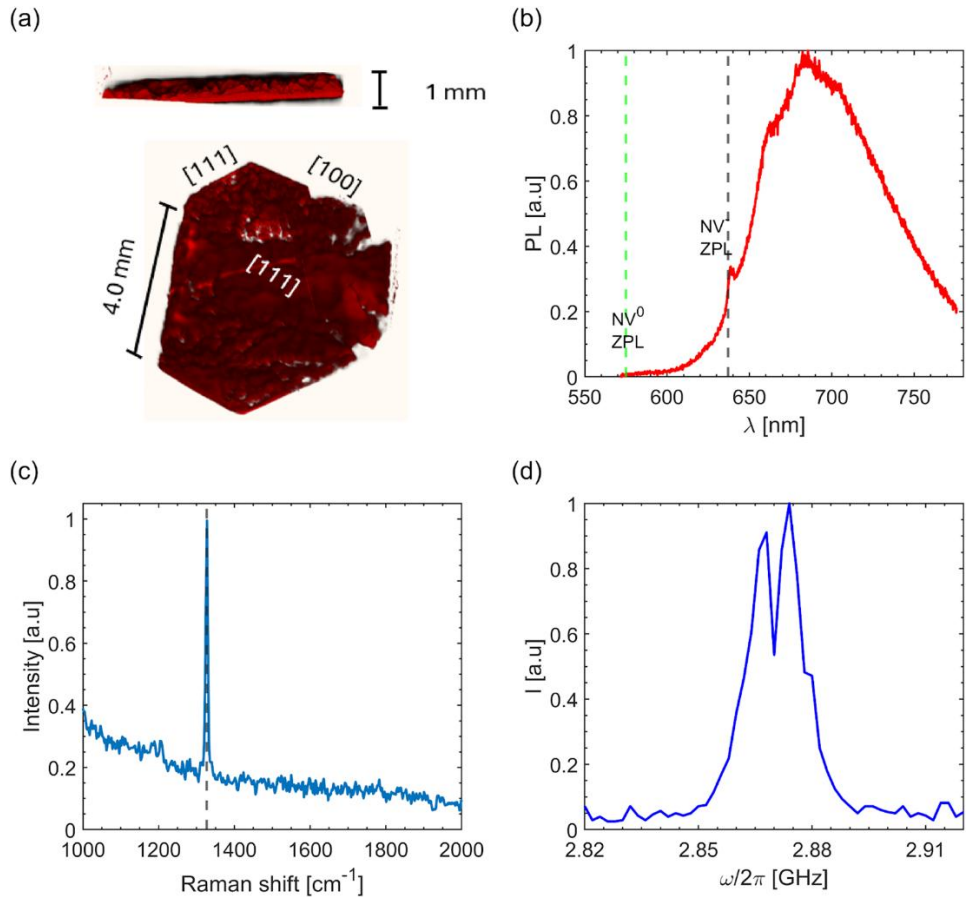

**Figure 1.** (a) 3D scan of the diamond sample in a confocal microscope. The diamond sample with [111] face orientation and two edges of 111 and 100, the sample length is 4.0 mm with a thickness of 1 mm. (b) Photoluminescence spectrum of the crystal with NV centers under excitation with a 552 nm laser. Two dashed lines indicate the zero-phonon lines of NV$^-$ and NV$^0$ at 637 nm and 575 nm, respectively. (c) Raman spectrum to reveal the peak at 1330 cm$^{-1}$ with no evidence of non-diamond carbon (i.e., G band 1500–1580 cm$^{-1}$)[27] or NV$^0$ (1406 cm$^{-1}$)[28] peaks. (d) ODMR spectrum with no applied magnetic field to reveal a resonance peak at 2.87 GHz. The peak splitting of 10 MHz is attributed to strain.

The Density Functional Theory analysis was then used to further support and quantify the experimental observations, explicitly calculating the nonlinear susceptibility introduced by the NV centers. The main DFT results schematics of NV in diamond are shown in **Figure 2**. Figure 2a shows the diamond supercell consisting of 2 x 2 x 2 unit cells with a single NV center marked in blue. The carbon-carbon bond lengths from the simulations are 7.156 and 1.56 Å respectively. This compares well to the literature with experimental values from the literature of 7.165 and 1.55 Å. The red arrow indicates the calculated dipole moment on the nitrogen atom. Figure 2b shows the overall charge distribution in the diamond super cell, with isosurfaces equal to 1.7. Figure 2c and Figure 2d show the LUMO and HOMO orbitals.

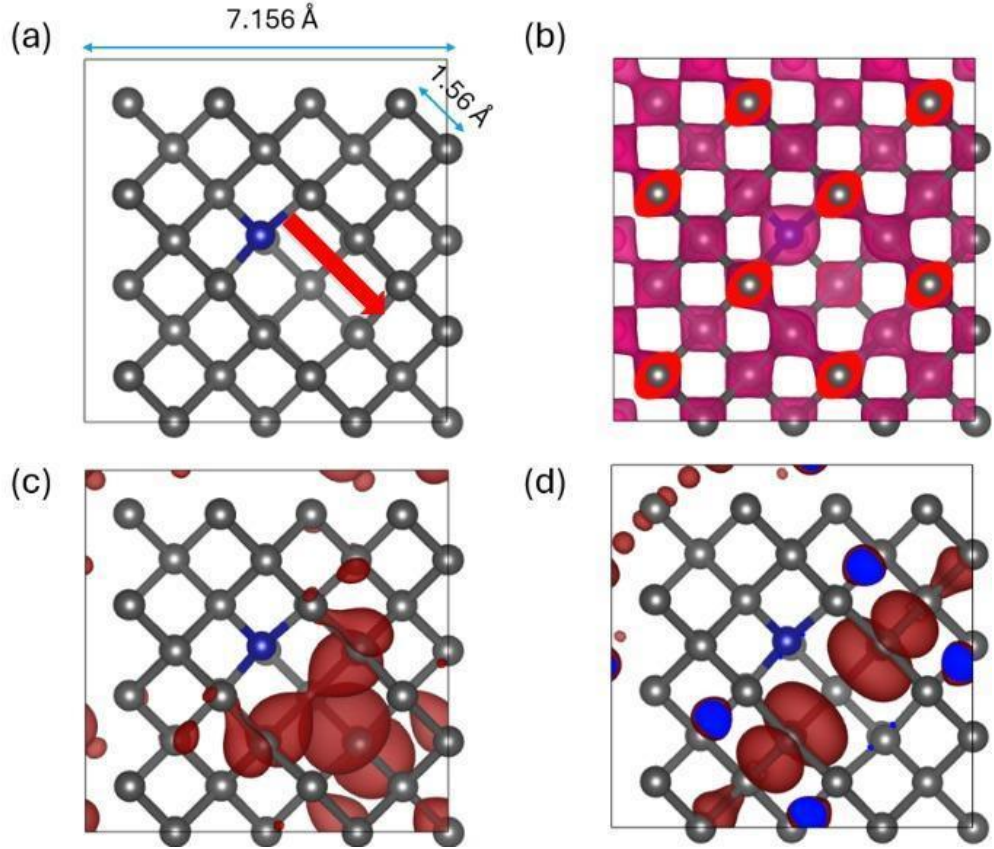

**Figure 2.** Optimized diamond cell according to DFT calculations. (a) A diamond super cell of 2 x 2 x 2 unit cells containing a single NV center. The carbon-carbon bond length and super cell length are 1.56 and 7.156 Å, respectively. According to experimental results, the carbon-carbon bond length and super cell length are 1.55 and 7.134 Å, respectively. The red arrow indicates the calculated dipole direction in the supercell (b) charge distribution in the diamond super cell, with isosurfaces equals to 1.7 (c) LUMO and (d) HOMO orbitals.

Explicit calculations of the NV wavefunctions (HOMO and LUMO orbitals shown in Figures 2c and 2d) indicate significant electronic localization around the defect site. This localization is directly responsible for the induced dipole moment and consequently the second-order nonlinear susceptibility $\chi^{(2)}$ observed experimentally. Such localized electronic states created by NV centers are critical in enhancing the material's optical rectification capability.

The linear and nonlinear properties of the crystal were examined in the NIR spectral range and are shown in **Figure 3**. Figure 3a shows the general orientation of the crystal, the crystal faces and the orientation axes of the pump polarization and generates THz pulse. The linear response of the crystal is shown in Figure 3b. The total transmission in the NIR range is in the region of 55 % from 900 to 1600 nm. The crystal was then pumped with NIR ultrashort pulses (1300 nm, 50 fs, 280 GW/cm²) and the generated THz signal was examined in transmission through the crystal using a THz time domain spectroscopy setup (see Methods for a full experimental description). Figure 3c shows the time domain THz electric field trace after pumping with a NIR ultrashort pulse, demonstrating a near single cycle waveform linearly polarized along the y axis, perpendicular to the 111 plane. Figure 3d shows the corresponding spectral information of the time trace, with a peak frequency around 1 THz. The maximum measured frequency was approximately 4.5 THz. The broadband nature of the emission (up to ~ 4.5 THz) demonstrates effective

~~approximately~~ phase matching between the pump and generated THz waves, enabled by the diamond's refractive index with a relatively flat dispersion from NIR to THz frequencies. Figure 3e shows the measured THz field intensity as a function of the pump intensity. The THz intensity increases with a second-order trend with respect to the pump intensity, as expected for such a nonlinear process. No saturation of the generated THz intensity was observed for the range of pump intensities used. Also, the effect of the pump linear polarization angle on the THz field strength was examined, shown in Figure 3f. In general, THz emission is observed along the y polarization direction for all input pump polarization angles, with some minor deviations in the overall field strength depending on the pump polarization angle. The observed insensitivity to pump polarization confirms that the dominant contribution to the nonlinear response comes from intrinsic dipoles associated with NV defects rather than any bulk or interface-driven anisotropy. Finally, the THz electric field strength was compared to a standard ZnTe crystal (d = 100 µm), for both crystals at a wavelength of 1300 nm and a pump intensity of 280 GW/cm². The resulting peak THz electric field emitted from the ZnTe was approximately 15 times more than the diamond-NV sample, as shown in the inset of Figure 3c This lower emission compared to ZnTe may be due to several crystal properties which may lead to a non-ideal environment for optical rectification of the pumping pulse. Firstly, the density of NV's and the fabrication method leads to transmission through the crystal of approximately 55 % in the NIR, potentially reducing the NV density may result in higher pump transmission and lower reabsorption of the generated THz pulse. Reducing NV density or optimizing their spatial distribution could indeed enhance THz emission by balancing the competing effects of pump absorption and nonlinear polarization generation. In the current configuration, it is likely that only emission from the end section of the crystal is contributing to the buildup of the THz pulse. Further investigations into the NV fabrication method and conditions may lead to better pump-THz phase matching, resulting in increased THz field emission.

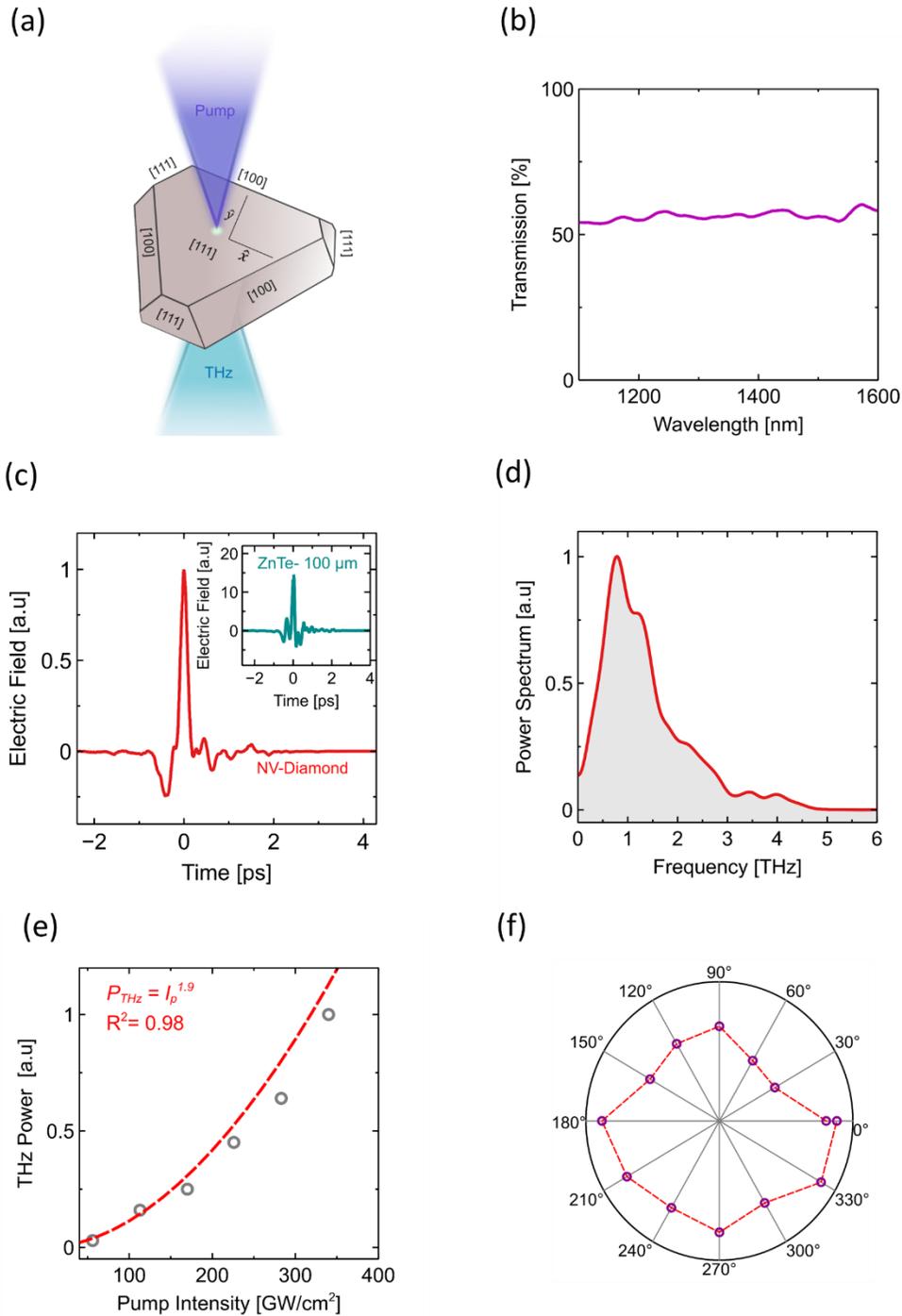

**Figure 3.** (a) Schematic of the diamond crystal showing the relative orientations of the crystal face, the pump incidence direction, and the THz emission detection. The pump is linearly polarized in the x-y plane and adjusted to study its impact on the THz emission. For panels (b-e), the pump is x-polarized. (b) Linear optical transmission of the diamond crystal in the NIR range, demonstrating typical values around 55 %. (c) Time-domain electric field traces for a 1300 nm pump with an input intensity of 280 GW/cm². Inset: Reference time-domain electric field emitted from a 100 μm ZnTe crystal under the same experimental conditions. (d) THz pulse spectrum from the time trace shown in (c), demonstrating spectral content up to 4.5 THz. (e) THz power as a function of the pump intensity. The fit

corresponds to the power of 1.9. (f) Peak THz field as a function of the linear input polarization angle, where 0° and 90° correspond to the x and y axes shown in (a).

Despite its experimentally lower THz yield $\sim 1/15$ that of ZnTe under comparable pumping, NV-doped diamond holds substantial promise once structural considerations of the hyperpolarizability $\beta$, effective $\chi_{eff}^{(2)}$, and alignment are enhanced. Assuming one NV center, the first hyperpolarizability $\beta$ is estimated on the order of $10^{-37} - 10^{-36} C \cdot m^2 \cdot V^{-2}$ considering the local field corrections, collectively producing an effective $\chi_{eff}^{(2)}$ in the range of $10^{-13} - 10^{-12} \, mV^{-1}$, given a moderate NV density $\sim 20 \, ppm$ and partial orientation. These values translate to an observed $d_{eff} \approx$ 1-5 pm/V, which can reach $5 - 10 \, pm/V$ with improved doping ($10^{17} - 10^{18} \, cm^{-3}$) and minimized $NV^0$ losses, placing NV-diamond in line with moderate efficiency THz emitters like GaP while exceeding them in thermal and mechanical robustness. ZnTe is limited to $\sim 3 \, THz$, while organic crystals exhibit large $\chi^{(2)}$ are fragile and often limited by phonon absorption. In contrast, NV-diamond remains broadband phase-matched owing to a refractive index difference $\Delta n \approx 0$ from 800 nm to $\sim 5 \, THz$. Furthermore, the crystal tolerates intense femtosecond beams due to diamonds' high damage threshold and retains quantum spin functionalities that might be exploited for integrated THz sensing or spin-lattice coupling. Current absorption and reflection consisting of $\sim 45\%$ the pump and suboptimal NV alignment, yielding only partial $\theta$-dependent coherence, restrict the fully realized $\chi_{eff}^{(2)}$. Further development of the diamond sample may be needed to enhance the THz emission, including optimization of the sample thickness, NV density, and the careful orientation of the detect axes. Taking these into account, NV-diamond can potentially outperform conventional THz emitters, leveraging the unique optical properties of the diamond lattice.

We have demonstrated for the first time broadband THz pulse emission from bulk diamond crystals through second-order nonlinear optical rectification induced by symmetry-breaking NV centers. Embedded NV centers act as localized dipole sources, effectively reducing the intrinsic centrosymmetric symmetry of the diamond, enabling otherwise even-order nonlinear optical processes to occur. Our experimental results show the emission of single-cycle THz pulses with spectral content exceeding 4.5 THz, driven by femtosecond NIR pulses without external bias fields or structural modifications. DFT calculations confirm the presence of localized electronic states around the NV centers, directly linking these states to the experimentally observed nonlinear susceptibility. Our rigorous quantitative analysis estimates the effective second-order susceptibility $\chi^{(2)}$ to be approximately $0.02 - 2 \, pmV^{-1}$, 15-fold lower than traditional nonlinear crystals like ZnTe, yet sufficient to yield measurable THz emission due to diamond's advantageous refractive index dispersion and high damage threshold. The NV hyperpolarizability ($\beta = 10^{-37} Cm^2 N^{-2}$) was derived, revealing that each NV center contributes substantially to the overall nonlinearity despite their individually modest nonlinear response. Current efficiency limitations primarily arise from absorption losses and phase mismatch due to NV density and distribution. However, our theoretical modeling suggests optimization, by refining NV implantation methods and defect alignment, we expect significant improvements in pump transmission, coherence length, and effective nonlinear interaction length. The intrinsic robustness, high thermal conductivity, broad transparency, and phase-matching capability position NV-diamond as a highly promising platform for THz generation, potentially overcoming limitations faced by traditional materials such as ZnTe or organic nonlinear crystals. Our findings open a promising new pathway toward compact, efficient, and broadband THz sources based on quantum-engineered diamond crystals.

## Supporting Information

The Supplementary Information contains extended methodological details, including diamond synthesis, NV center activation, and the experimental setups for optical imaging, Raman spectroscopy, ODMR, and THz time-domain spectroscopy. It also provides in-depth quantum calculations using density functional theory (DFT) to analyze the electronic structure and nonlinear optical properties of NV-doped diamond. Additionally, it includes a detailed discussion of the nonlinear optical framework, phase-matching conditions, and theoretical modeling of THz emission. Supporting data such as dielectric tensor components and supplementary figures further substantiate the experimental findings presented in the main text.

## Acknowledgments

M.A. would like to thank the Israeli Ministry of Innovation, Science and Technology for their support. This project was funded by the Israel Science Foundation (ISF), Grant No. 1357/21 (A.B.) It was also partially supported by the Technion's Hellen Diller Quantum Center and its Russell Berrie Nanotechnology Institute (A.B.).

## Conflict of Interest

The authors declare no conflict of interest.

**TOC Figure**

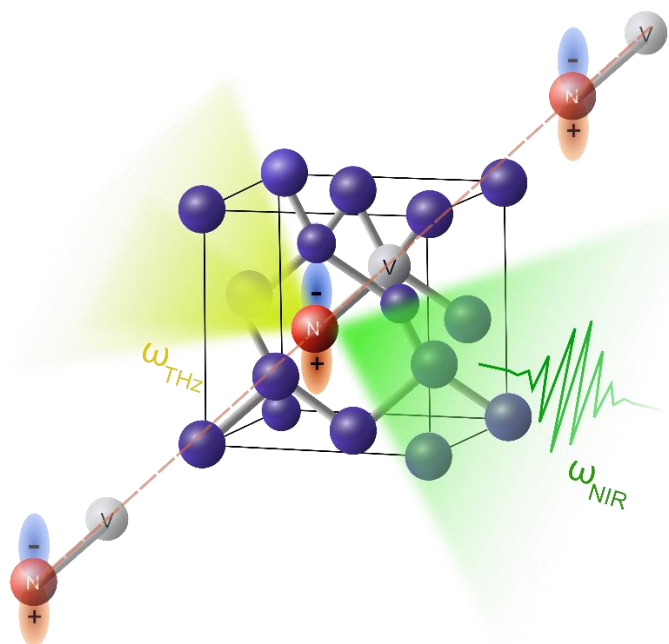